\documentstyle[12pt]{article}
\newcommand{\be}{\begin{equation}}
\newcommand{\ee}{\end{equation}}
\begin{document}
\title{\huge Body Motion In a Resistive Medium at Temperature $T$.
\author{{\bf M. I.  Molina}
\vspace{0.4 cm}
\and
\and
Facultad de Ciencias, Departamento de F\'{\i}sica, Universidad de Chile\\
Casilla 653, Las Palmeras 3425, Santiago, Chile.\\
mmolina@abello.dic.uchile.cl}}
\date{}
\maketitle
\baselineskip 20 pt
\vspace{0.0in}
\begin{center}
{\bf Abstract}
\end{center}
We consider a macroscopic body propagating in a one-dimensional
resistive medium, consisting of an ideal gas at temperature $T$.
For a whole family of collisions with varying degree of inelasticity,
we find an exact expression for the effective force on the moving
body as a function of the body's speed and the value of the restitution
coefficient. At low  and high speeds it reduces to the well-known
Stoke's  and Newton's law, respectively.
\vspace{0.2cm}

\noindent
Se considera un cuerpo macrosc\'{o}pico propag\'{a}ndose en un
medio resistivo unidimensional, consistente de un gas ideal a
temperature $T$. Para toda una familia de colisiones
con diferente grado de inelasticidad, hallamos una expresi\'{o}n
exacta para la fuerza efectiva sobre el cuerpo como funci\'{o}n de la
velocidad del cuerpo y del coeficiente de restituci\'{o}n.
A bajas y altas velocidades, se reduce a la conocida
ley de Stoke y Newton, respectivamente.\\
{\em Descriptores:\ roce viscoso, colisiones.}
\vspace{0.2cm}

\noindent
\centerline{\em PACS:\ 45.20.Dd , 45.50.Tn \ }

\newpage
\section{Introduction}
\noindent
When an object moves through a viscous medium, such as water or air,
it experiences a resistive drag force. For small objects such as dust
particles movingh at low speeds, this drag force is proportional to
the speed of the object. This is known as Stoke's law\cite{marion}.
For larger objects such as airplanes, skidivers and baseballs
moving at high speeds the drag force is approximately proportional to
the square of the speed\cite{parker}.
This limit is known as Newton's law. The general problem of
determining the exact dependence
of the drag force on the speed of an arbitrarily-shaped
object moving at any speed, defies any
closed form solution given its complex many-body character.
A complete solution would have to
take into account the detailed scattering process between
the body and the particles
composing the medium, the thermal properties of the medium, the presence of possible internal degrees
of freedom of the body and local turbulence effects, etc.
However, beneath all these complexities is basically
the transfer of momentum and energy between the body
and the medium particles.
Therefore, is instructive to consider simplified ``toy'' models where one can track in detail the momentum
and energy exchange between the body and its surrounding medium.
This is realized at the expense of simplifying other
factors such as the dimensionality of the system or the specific form of the interactions between
the body and the medium.
In this spirit, we present here an extension of a previous\cite{revmex1}, simplified  zero-temperature
one-dimensional model, where we now include finite temperature effects. This immediately brings into
the game a natural velocity scale not present in our previous model: the thermal speed.
We obtain the resistive drag force as a function of the
body's speed in {\bf closed form} and find that, when the speed of the body is smaller than the
thermal speed, the resistive force is {\em linearly} proportional to the speed of the body.
On the contrary, when the speed of the body is greater then the thermal speed, the
proportionality becomes {\em quadratic}.   

\section{The Model}

\noindent
Let us consider a (macroscopic) body of mass $M$ propagating in a
one-dimensional resistive medium modelled by an ideal gas in
thermodynamic equilibrium at temperature $T$, characterized
by a thermal speed $V_{T} \equiv \sqrt{k T /m}$, where $k$
is Boltzmann's constant and $m$ is the mass of a medium
particle (Fig. 1). We assume the body to be truly macroscopic,
like a baseball moving through air, or a falling rock.
In other words, $M \gg m$ which allows us to make the
following simplification:  During a medium particle-body
collision event, we will take the mass of the body to be
essentially infinite. In this approximation the body is
pictured as a massive, partially absorbing, moving
``wall'' colliding constantly with the medium
particles. A reasonable assumption, if one considers
that the mass ratio $m/M$ is of the order of $10^{-24}$
for a baseball moving through air. After each collision,
the speed of the body is essentially unchanged, so the
magnitude of the momentum transferred to the body is
\[
\Delta p \approx (1 + \epsilon) m | V - v |
\]
where $V$ is the speed of the body, $v$ the speed of
the medium particle and $\epsilon$ is the restitution
coefficient for the body-particle collision. Thus,
when $\epsilon = 1$ we have a completely elastic collision,
where the magnitude of the relative
body-particle velocity is conserved, while at $\epsilon = 0$,
we have the case of a completely inelastic collision,
where the particle is ``absorbed'' by the body after
colliding. We also work in a quasi-continuum approximation
where an element of length $dx$ while
``small'', will contain a large number of medium
particles. 

Initially the body is given a speed $V_{0}$ (say, to the right),
and we observe the system at a later time $t$,
when the speed of the body is $V$. During the next time
interval $d t$, the body will collide with particles coming
from its left and right side. On the left side, only those particles
that have speeds $v > V$ and are located closer than $(v - V) dt$
will collide with the body. The number of such particles is
$dn_{L} = \rho dn(v) \Theta (v - V) (v - V) dt$.
In a similar manner, the number of particles to the right of the body that
will collide with the body during the interval $dt$ is
$dn_{R} = \rho dn(v) \Theta(V -v) (V-v) dt$.
Here $\rho$ is the particle number density,
$\Theta(x)$ is the step function ($\Theta(x) = 1$, for $x >0$,
zero otherwise) and  $dn(v)$ is the number
of particles that have speeds in the interval $[v, v + dv]$:
$dn(v) = g(v) dv$ where $g(v)$ is the
thermal speed distribution, given by
\be
g(v) = (1/\sqrt{2 \pi}) {1\over{V_{T}}} \exp[ -{1\over{2}}\left( v/V_{T}\right)^2].\label{eq:P(v)}
\ee
The transfer of momentum per unit time coming from the medium to the
{\em left} of the body, due to particles with speed in the interval
$[v, v + dv]$ would then be:
\be
{d P\over{dt}} = (1 + \epsilon) m (v - V) g(v) \Theta(v - V) (v - V) dv
\ee
By integrating over all speeds, we obtain the
average effective force on the body from the left side:
\be
F_{\mbox{\ left}} = \int_{V}^{\infty} \rho m (1 + \epsilon) (v - V)^{2} g(v) dv.\label{eq:Fl}
\ee
In a similar manner, the transfer of momentum per unit time coming 
from the medium to the {\em right} of the body,
due to particles with speed in the interval
$[v, v + dv]$ is:
\be
{d P\over{dt}} = -(1 + \epsilon) m (V - v) g(v) \Theta(V - v) (V - v) dv
\ee
which implies that the average force on the body from the right side is
\be
F_{\mbox{\ right}} = \int_{-\infty}^{V} \rho (1 + \epsilon) (V - v)^{2} g(v) dv\label{eq:Fr}
\ee
The net average force $F$ on the body, along the direction of its initial velocity,
is given by the difference between Eq.(\ref{eq:Fl})
and (\ref{eq:Fr}):
\[
F  = - m \rho (1 + \epsilon) \left[\ \int_{-\infty}^{V} \rho (1 + \epsilon) (V - v)^{2} g(v) dv -
\int_{V}^{\infty} \rho m (1 + \epsilon) (v - V)^{2} g(v) dv\ \right]
\]
By inserting expression (\ref{eq:P(v)}) for $g(v)$ and carrying out the integrations, we obtain:
\be
F = - m \rho (1 + \epsilon) V_{T}^{2} \left\{ {\sqrt{2\over{\pi}}} \left( {V\over{V_{T}}}\right)\ \exp\left[-{1\over{2}}\left( {V\over{V_{T}}}\right)^2\right] +
\left( 1 + \left( {V\over{V_{T}}} \right)^2 \right)\ {\mbox{Erf}}\left( {V\over{\sqrt{2} V_{T}}}\right) \right\}\label{eq:F}
\ee
This rather complex-looking expression is a bit deceiving since
it depends on  negative exponentials of $(V/V_{T})^{2}$ which
makes it very sensitive to whether $V/V_{T} <1$ or $V/V_{T} > 1$.
In other words, we expect two, well-defined behavior regimes, with a small
crossover region near $V/V_{T} \approx 1$.

\section{Results and discussion}

\noindent
As can be clearly seen from (\ref{eq:F}),
the degree of inelasticity plays only a minor role, renormalizing
the number density of the medium.
Figure 2 is a log-log plot of the effective average force on the
body as a function of the speed of
the body, Eq.(\ref{eq:F}). As anticipated above, we note that
except for a small vicinity around
$V = V_{T}$,
it consists of basically two straight lines with slopes of one and two,
respectively. That is, at
speeds smaller than the thermal speed $V_{T}$,
the resistive force is proportional
to the body's speed (Stoke's law); while for body's speeds greater than $V_{T}$, the resistive force
becomes quadratic on the body's speed (Newton's law).
These limits are easy to derive from Eq.(\ref{eq:F}):
For $V \ll V_{T}$, ${\mbox{Erf}}(V/\sqrt{2} V_{T})
\approx \sqrt{2/\pi} (V/\sqrt{2} V_{T})$ and
$\exp[-(1/2)(V/V_{T})^2] \approx 1$, which implies:
\be
F \approx -\sqrt{8/\pi} m \rho (1 + \epsilon) V_{T} V = -\rho (1 + \epsilon)
\sqrt{8 m k T/\pi}\ V\hspace{1cm}\mbox{$V \ll V_{T}$}.
\ee
On the other hand, when $V \gg V_{T}$, ${\mbox{Erf}}(V/\sqrt{2} V_{T}) \approx 1$ and
$\exp[-(1/2)(V/V_{T})^2] \approx 0$. Thus, in this case one has:
\be
F \approx -m \rho (1 + \epsilon) V^{2}\hspace{2cm}\mbox{$V \gg V_{T}$}.
\ee
 
Let us now consider the issue of the stopping distance. For a medium
at a finite temperature, the speed of the body decreases
(on a macroscopic scale) as it moves through the medium and will eventually become
smaller than the thermal speed. At that point, the resistive force becomes
proportional to the speed, $ F = -\beta\ V$.
A simple integration then leads to an
exponential decrease on $V$ and therefore, a {\em finite} stopping distance. 
If the medium is at zero temperature however, the resistive force is
always quadratic with speed $ F = \gamma V^{2}$ and, in that case, it
can be easily proved that the stopping distance diverges logarithmically
with time\cite{revmex1}.

In summary, we have examined a simplified model of a macroscopic object
propagating in a resistive one-dimensional medium modelled as
an ideal gas at temperature $T$. For general inelastic collisions
between the body and the medium particles, characterized by a
restitution coefficient $\epsilon$, $0 \le \epsilon \le 1$,
we have arrived at a closed-form solution for the resistive
force in terms of the speed of the body.
Below the thermal speed, this force is essentially linear
in the body'd speed, while above thermal speed, the dependence
becomes quadratic.

\newpage

\centerline{\bf Figure Captions}
\vspace{2cm}  

\noindent
{\bf FIG 1:}\ \ A macroscopic body of mass $M$ propagating inside a one-dimensional
resistive medium composed of an ideal gas of particles of mass $m$, with $m \ll M$, in thermal equilibrium
at temperature $T$. The body undergoes partially elastic collisions with the medium particles
with restitution coefficient $\epsilon$.
\vspace{1cm}

\noindent
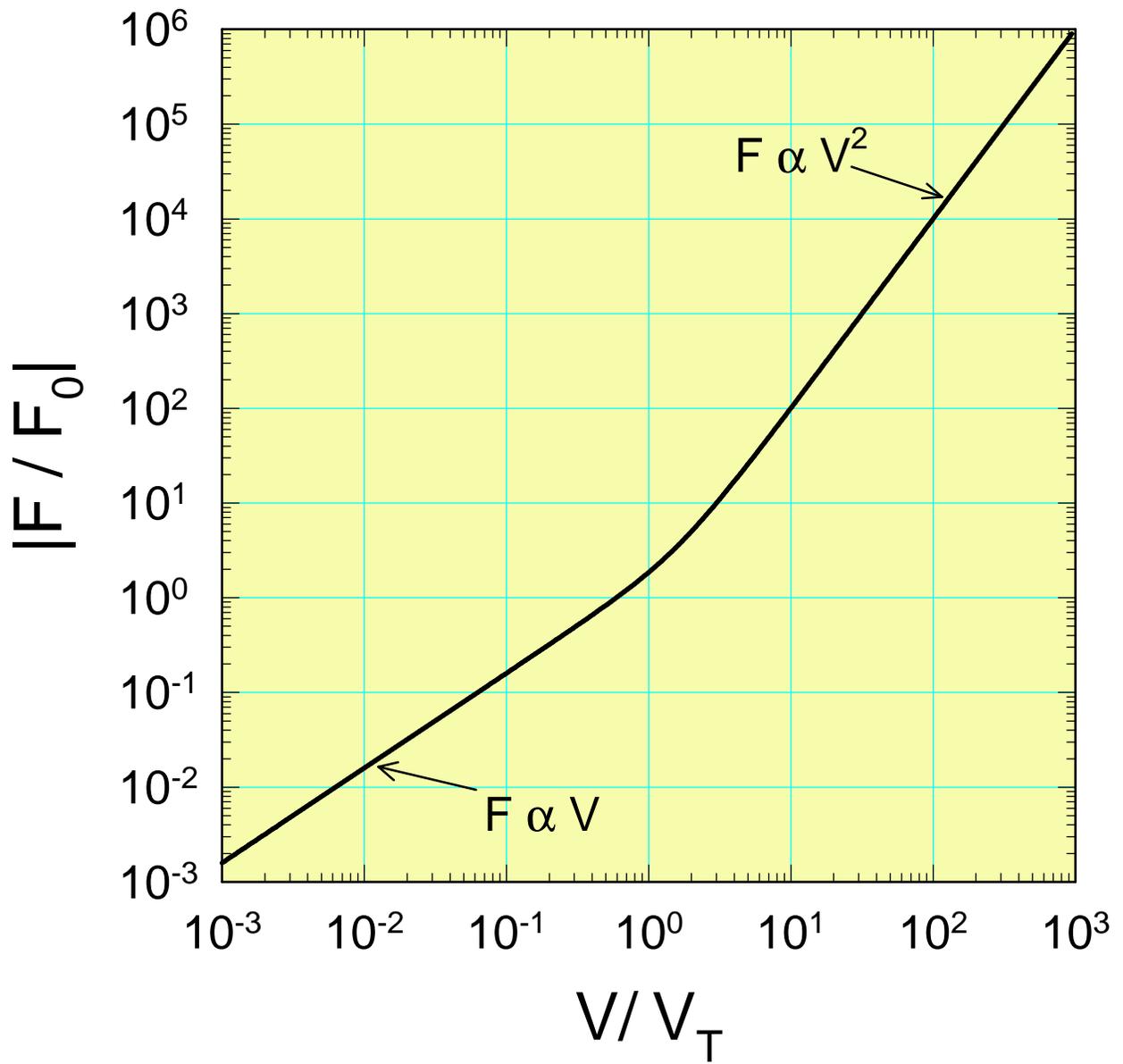
{\bf Fig. 2:}\ \ Effective average force on the macroscopic body,
as a function of the body's speed. For speeds smaller (higher) than the
thermal speed, the dependence is essentially linear (quadratic).
The crossover region is confined to a small vicinity around $V_{T}$.
( $F_{0} \equiv m \rho (1 + \epsilon) V_{T}^{2}$ ). 
\vspace{1cm}


\newpage

\begin{thebibliography}{99}
\vspace{2cm}
\bibitem{marion}
J. B. Marion and S. T. Thornton, {\em Classical Dynamics of Particles and Systems}
(Saunders College Publishing, Philadelphia, 1995), 4th. ed., pp. 60--71.

\bibitem{parker}
G. W. Parker, Am. J. Phys., {\bf 45}, 606 (1977)

\bibitem{revmex1}
M. I. Molina, Rev. Mex. Phys. {\bf 47}, 201 (2001).


\end{thebibliography}
\end{document}